# Visualization of the Citation Impact Environments of Scientific Journals: An online mapping exercise

Loet Leydesdorff

Amsterdam School of Communications Research (ASCoR), University of Amsterdam, Kloveniersburgwal 48, 1012 CX  Amsterdam, The Netherlands

loet@leydesdorff.net; http://www.leydesdorff.net

**Abstract**

Aggregated journal-journal citation networks based on the *Journal Citation Reports 2004* of the *Science Citation Index* (5968 journals) and the *Social Science Citation Index* (1712 journals) are made accessible from the perspective of any of these journals. A vector-space model is used for normalization, and the results are brought online at http://www.leydesdorff.net/jcr04 as input-files for the visualization program Pajek. The user is thus able to analyze the citation environment in terms of links and graphs. Furthermore, the local impact of a journal is defined as its share of the total citations in the specific journal's citation environments; the vertical size of the nodes is varied proportionally to this citation impact. The horizontal size of each node can be used to provide the same information after correction for within-journal (self-)citations. In the "citing" environment, the equivalents of this measure can be considered as a citation activity index which maps how the relevant journal environment is perceived by the collective of authors of a given journal. As a policy application, the mechanism of interdisciplinary developments among the sciences is elaborated for the case of nanotechnology journals.

*I've often stressed the importance of limiting comparisons between journals to those in the same field.*
Garfield, 1980, p. 1A

## 1. Introduction

On the basis of an experimental version of the *Science Citation Index* in 1961, Derek de Solla Price (1965, p. 515) formulated a program for mapping the sciences in terms of aggregated journal-journal citation structures as follows:

> The total research front of science has never, however, been a single row of knitting. It is, instead, divided by dropped stitches into quite small segments and strips. From a study of the citations of journals by journals I come to the conclusion that most of these strips correspond to the work of, at most a few hundred men at any one time. Such strips represent objectively defined subjects whose description may vary materially from year to year but which remain otherwise an intellectual whole. If one would work out the nature of such strips, it might lead to a method for delineating the topography of current scientific literature. […] Journal citations provide the most readily available data for a test of such methods.

Price had been fascinated with journals and their exponential growth in size and numbers ever since his study of the *Philosophical Transactions of the Royal Society of London* from its very beginning in 1665 (Price, 1951; 1961; 1978). Can the aggregated citation relations among journals be used to study clusters of journals as representations of the intellectual organization of the sciences? I have addressed this question in my research over the past twenty years because three theoretically important problems could be addressed if the intellectual organization of the sciences could be operationalized using journal structures:

1. In science studies this operationalization of the intellectual organization of knowledge in terms of texts (journals) as different from the social organization of the sciences in terms of institutions and people would enable us to explain the scientific enterprise as a result of these two interacting and potentially coevolving dimensions (Whitley, 1984; Leydesdorff, 1998).

2. In science policy analysis, the question of whether a baseline can be constructed for measuring the effectivity of political interventions was raised by Studer and Chubin (1980, p. 269; cf. Leydesdorff & Van der Schaar, 1987; Leydesdorff *et al.*, 1994). Van den Daele *et al.* (1979) already distinguished between parametric steering in terms of more institutional activities due to increased funding versus the relative autonomy and potential self-organization of scientific communication into specialties and disciplinary structures (Collins, 1985; Martin & Irvine, 1985; Leydesdorff, 1995).

3. Impact factors of journals are defined with reference to averages across the sciences (Garfield, 1979), while important parameters of intellectual organization like publication and citation frequencies can be expected to vary among disciplines (Price, 1970). Publication practices across disciplinary divides are virtually incomparable (e.g., Nederhof *et al.*, 1989; Van Gigch, 2002a, 2002b). The impact factor can be considered as a global measure which does not take into account the intellectual structure present in the database.

The alternative of a more finely-grained measure of impact like a local impact factor (Hirst, 1978), however, presumes either the possibility of a robust classification of the journals (Pinski & Narin, 1976) or it has to be based on another unit of analysis, e.g., the individual paper (Moed, 2005). Relations among individual papers can be mapped using co-citation or co-word analysis (Callon, 1986; Chen, 2003; Small, 1999; Garfield *et al.*, 2003). These maps enable us to follow historical developments *within* fields of science, but the identification of these quasi-objectified structures in terms of their disciplinary organization cannot be provided by the maps themselves (Rip, 1997). The external yardsticks remain the journals and the journal structures (Leydesdorff, 1987).

Can a robust way be found to delineate the database in terms of specialties and disciplines? After a long series of attempts to develop this methodology (e.g., Leydesdorff, 2002, 2003, 2004a, 2004b; Leydesdorff & Cozzens, 1993), I have come to the conclusion that this is not a viable project for both theoretical and empirical reasons. Empirically, the top-down decomposition and the bottom-up agglomeration can be distinguished (Leydesdorff, 2004b). The top-down decomposition has become possible recently since one is able to load the entire citation matrix into memory (Leydesdorff, forthcoming). The bottom-up aggregation remains very sensitive to the point(s) of entrance and other parameter choices because the multi-dimensionality of the journal space may bring together in a latent dimension what seems far apart in the dimensions under study. The problem finds its origin in the fuzziness of the sets: different sets are partial subsets of one another (Bradford, 1934; Garfield, 1979; Pudovkin & Garfield, 2002; Leydesdorff & Bensman, forthcoming).

The aggregated journal-journal citations provided by the *Journal Citation Reports* of the *(Social) Science Citation Index* can be considered as a huge matrix of cited and citing journals, respectively. The matrix is asymmetrical and overwhelmingly empty. Scientific journals tend to cite one another in dense clusters which represent specialties. However, some (e.g. interdisciplinary) journals cite and are cited across different fields (Narin *et al.*, 1972). This is well-known of *Science* and *Nature* at the top of the hierarchy, but there are also hierarchies spanning fields at lower levels (Doreian 1986; Doreian & Fararo, 1985). For example, the journals of American professional associations may function as elite institutions across cognitive delineations among specialties (Bensman, 1996). While the majority of the journals remain embedded in one or more specialized publication and citation structures, the matrix thus is *nearly* decomposable (Simon, 1973).

Consequently, the decomposition remains sensitive to the choices of the various parameters involved, such as the seed journal(s) for collecting a citation environment, the threshold levels, similarity criteria, and the clustering algorithm. In other words, the vectors of the journal distribution span a multi-dimensional space in which clouds can be distinguished, but the delineation of these clouds at the edges remains fuzzy (Bensman, 2001) and varies with the perspectives chosen by the analyst (Leydesdorff & Cozzens, 1993; McKain, 1991). Particularly, if one wishes to construct a baseline against which to measure change, the distinctions among variations, measurement errors, auto-correlations in the data, and structural change may become too uncertain to be meaningful (Leydesdorff, 1991; 2002).

While in previous mappings the search focused mainly on a parsimonious representation among the many possible ones (e.g., by using rotation of the main dimensions as in factor analysis or multi-dimensional scaling), the possibility to provide journal maps online using visualization techniques from social network analysis has changed the situation dramatically (Otte & Rousseau, 2002). If it is deemed no longer feasable to provide an objectified representation at the cluster level, *one might leave the choice of the entrance journal, the choice of the clustering, and therewith the perspective to the end-user*. Furthermore, users may wish to inspect the relevant citation environments in both the cited and citing dimensions as two different perspectives on a journal's position. Tijssen *et al.* (1997) combined these two perspectives into a single representation using quasi-correspondence analysis, but there are substantive reasons to distinguish between "cited" as impact and "citing" as behaviour. Zhou & Leydesdorff (2005), for example, found that leading journals in China sometimes cite exclusively from the international literature, but are cited mainly by other journals at the national level. Thus, cited and citing aggregates can inform us also about hierarchies among journals.

The practical applications of a visualization and quantification of the citation impact environments of journals are numerous. Librarians, for example, can use this information to improve the quality of their collections or compose a list of core journals relevant to their specific needs (Hirst, 1978). Prospective authors may be interested in neighbouring discourses and how these are relevant for their publication and citation profiles. The maps represent the subject structure in terms of the positions of the nodes and the links between them. The links allow users to detect the clusters in the graph either visually or by using more sophisticated tools like graph-theoretical algorithms (e.g., Bollen *et al.*, 2005). The size of the nodes can be used to represent the percentage of the citations within a specific citation environment either including or excluding within-journal (self-)citations.

## 2. Methods

This project developed in two stages. First, all journals of the *Journal Citation Report 2003* were mapped in terms of the cosines among the vectors of the journals in the environments of each journal under study. (I will use the words "seed journal" for this entrance point in the remainder of the article.) This was done for both the *Science Citation Index* (5907 journals) and the *Social Science Citation Index* (1710 journals). The relevant environment for each subsequent journal was determined by including all journals which cite or are cited by the journal under study to the extent of one percent of its citation rate in the respective dimension (He & Pao, 1986; Leydesdorff, 1986). This generates sets on the order of 10-50 journals. For each set, a citation (transaction) matrix can be composed.

As the similarity measure between the distributions for the various journals included in a citation environment, I use the cosine between the two vectors or, in other words, the geometrical mean (Salton & McGill, 1983).[1] Unlike the Pearson correlation coefficient, the cosine does not normalize for the arithmetic mean (Jones & Furnas, 1987). This has advantages in the case of sparse matrices (Ahlgren *et al.*, 2003). For the purpose of the visualization, it is convenient that the cosine provides us with positive values only, while one expects also negative values in a Pearson correlation matrix. While the Pearson correlation coefficient remains the statistical instrument for finding the eigenvectors of the network or for inferential statistics (Bensman, 2004), the cosine seems an appropriate

---

[1] The cosine of the angle enclosed between two vectors *x* and *y* is defined as follows:

$$\text{Cosine}(x,y) = \frac{\sum_{i=1}^{n} x_i y_i}{\sqrt{\sum_{i=1}^{n} x_i^2} \sqrt{\sum_{i=1}^{n} y_i^2}} = \frac{\sum_{i=1}^{n} x_i y_i}{\sqrt{(\sum_{i=1}^{n} x_i^2)*(\sum_{i=1}^{n} y_i^2)}}$$

measure for mapping the vector-space (Almind & Ingwersen, 1997; Leydesdorff & Vaughan, forthcoming).

For each journal two files were generated: one in the "cited" and another in the "citing" direction of the asymmetrical citation matrix. The (ASCII) text files can be read directly into Pajek. Pajek is a visualization program which is freely available for non-commercial usage at http://vlado.fmf.uni-lj.si/pub/networks/pajek . Cosine values below 0.2 were suppressed in order to enhance the interpretatibility of the visualizations. Because the cosine has values only between zero and one, low values of the cosine indicate negative correlations.

Within Pajek the user can choose a variable width for the lines in the network and colours or grey shades for nodes and links, respectively. The nodes can also be partitioned (and coloured) in accordance with their allocation into clusters using the various graph-analytical tools available within the program. Files can be exported in other formats for further processing, including the so-called DL-format which is used by programs like UCINET and NetDraw.

With hindsight, the analysis for 2003 can be considered as the development of a beta-version. The results were made available online in order to generate feedback from the community. Two further improvements were made in 2004. First, the citation environments are normalized in 2004 not with reference to the cited and the citing dimensions combined (using an OR statement), but independently. Thus, in 2003 each journal citing or being cited to the extent of more than one percent was drawn into the environment, while the environments are specific and potentially different for "being-cited" or "citing" in 2004. The two citation environments can be very different.

Secondly, the percentage of contributions to the citations—citing or cited, respectively—will be used to determine the size of each node. By distinguishing between the vertical size of a node and the horizontal one, a second parameter can be used to indicate this percentage after correction for within-journal (self-)citations (Price, 1981; Noma, 1982).

Thus, by inspecting the shape of the ellipses one is able to see how much a journal is dependent on an inner circle of authors citing one another. Note that within-journal citations can be both self-citations of authors and citations among authors publishing in the same journal.

The reasons to normalize in terms of numbers of citations as an indicator of local impact were two-fold. Initially, I considered a normalization of the numbers of citations divided by the number of publications—the c/p ratio—for each journal. However, the number of publications to take into account is ill-defined, and the citation window would be unlimited (while it is limited to two years in the case of the impact factor as defined by the Institute of Scientific Information (ISI)). In an extreme case, for example, a journal might have disappeared ($p = 0$) while it is still being cited, and the c/p ratio would therefore go to infinity. More importantly, however, in one of the rare validation studies, Bensman (forthcoming) compared (1) survey results of the chemistry department of Louisiana State University, and (2) a journal use study at the University of Illinois Chemistry Library, with (3) impact factors and (4) total cites for the same year (1993). The author found correlations between "total cites" and the appreciations by users which were significantly higher than the correlations of the latter with the impact factor. Bensman kindly made his data available to me. The number of publications was added so that c/p ratios could also be calculated.

Of course, this is not the place to publish Bensman's (forthcoming) results (Bensman, 2001; Bensman & Wilder, 1998). However, a factor analysis of these variables teaches us that two factors explain some 82% of the variance. The two indices which are normalized for the number of publications—the c/p ratio and the impact factor—correlate highly, while an orthogonal dimension is spanned by a factor which could perhaps be designated as the "prestige" of the journal in question. Prestige is not determined by average impact, but by the top range of a highly skewed distribution (Brewer *et al.*, 2001; Seglen, 1997). In other words, these results confirm Garfield's (1998) conjecture that: "The use of journal impact factors as surrogates for actual citation performance is to be avoided, if at all possible."

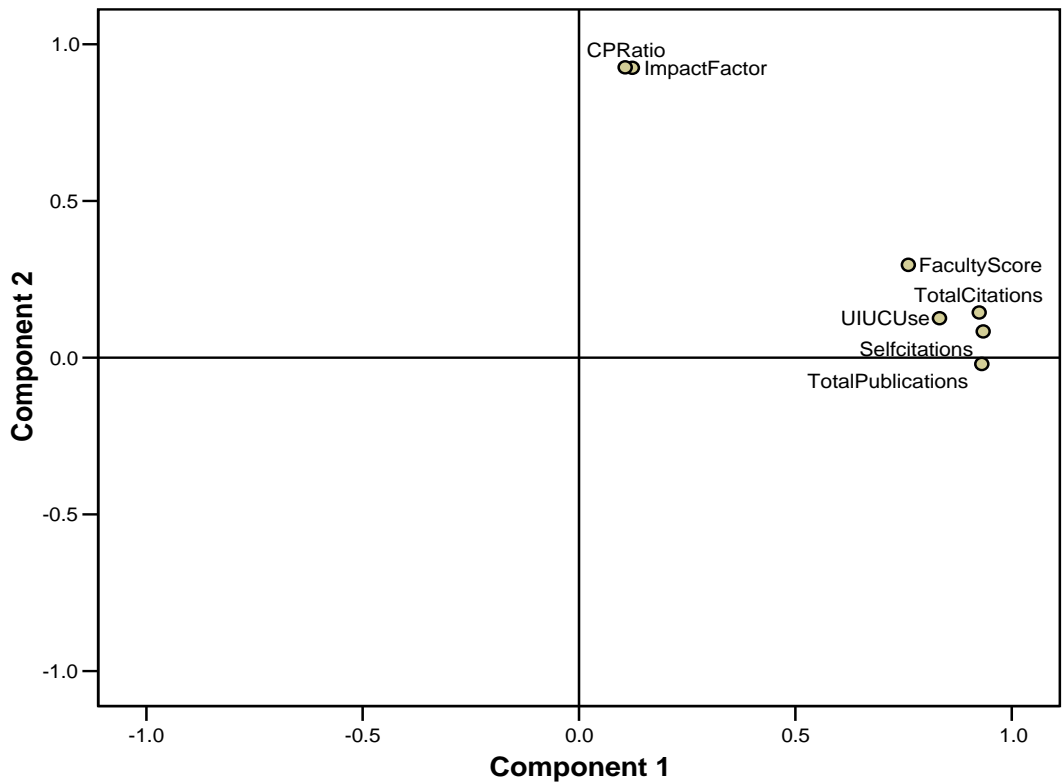

**Figure 1**: Component plot in rotated space (sources: JCR, 1993; Bensman, 2001; forthcoming; Bensman & Wilder, 1998).

In summary, all journals citing or being cited by the seed journal will be drawn into the local citation environment, respectively, but the tail of the distributions from the seed journal's perspective is discarded for delineating the environment. Thereafter, all values above one are used for the citation matrix (because the ISI suppresses single relations by summing them under the category "All others"). The grandsum of the consequent citation matrix $N \; (= \sum c_{ij})$ is used as the basis for the normalization of the citation contributions. Each journal contributes with its margin total $n_i \; (= \sum c_i)$ as a percentage of the grandsum. The value of the main diagonal element ($c_{ii}$) can additionally be used as a correction factor.

3. **Materials**

The data was harvested from the CD-Rom versions of the *Journal Citation Reports* for the respective years. The bottom values for "All others" combined were discarded, and all other data organized in a relational database management system so that citation matrices can be extracted from any chosen perspective. In principle, these perspectives can be based on choosing either a seed journal or a list of relevant journals, but we limit the analysis in this study to individual journals. A dedicated program was written to provide the citation matrices as input files for SPSS, UCINET, and Pajek. The focus is here on the Pajek files because these are the ones brought online. The discussion, however, is informed in terms of relevant *dimensions* (eigenvectors) by using factor analysis in SPSS (as will be demonstrated below in Table 2 when discussing the first results).

Table 1 provides descriptive statistics for the two databases (*Science Citation Index* and *Social Science Citation Index*) for the two years (2003 and 2004), respectively.

|  | *SCI 2003* | *SCI 2004* | *SoSCI 2003* | *SoSCI 2004* |
|---|---|---|---|---|
| number of source journals processed | 5714 | 5968 | 1708 | 1712 |
| source journals not processed 'citing' | 193 | 192 | 52 | 40 |
| unique journal-journal relations | 917,502 *2.91%* | 1,038,268 *3.01%* | 90,454 *3.20%* | 96,207 *3.36%* |
| sum of journal-journal relations | 17,604,594 | 18,943,827 | 902,808 | 966,619 |
| *average cell value* | *19.19* | *18.25* | *9.98* | *10.05* |
| total 'citing' | 23,953,246 | 25,798,965 | 2,714,493 | 2,909,219 |
| total 'cited' | 19,497,302 | 20,909,401 | 1,337,339 | 1,453,397 |
| within-journal citations' | 1,960,559 | 2,016,500 | 136,948 | 137,269 |

**Table 1**: Descriptive statistics of the JCRs 2003 and 2004, of the *Science Citation Index* and the *Social Science Citation Index*, respectively.

Note that the within-journal citations are on the order of 10% across the files, but as we shall see below this percentage differs considerably among journals and specialties. The density of the network is more than twice as high in the sciences when compared with the social sciences in terms of the average cell values. All matrices are extremely sparse: I added to the row with unique journal-journal relations (in italics) the percentage of cells with a value as compared with the total number of cells of the matrix, that is, the number of possible citation relations among journals in the set. This percentage is a bit higher for the social sciences because the networks are more spread.

Because of the copyright issues potentially involved in using the data, I bring only the normalized matrices of cosines and not the data matrices online. In previous studies (Leydesdorff, 2004a and b), I made the visualizations available as pictures, but the input files enable users to apply their own visualization techniques and clustering algorithms in a more flexible way. The visualizations in this study are based on using the algorithm of Kamada and Kawai (1989) as it is available in Pajek unless otherwise indicated.[2]

## 4. Results

The matrix of cosine values are organized using UCINET's so-called DL language in 2003 and the Pajek format in 2004. In both formats the files are plain text and they can be converted into each other using either program. The user may also wish to edit the files. As noted, the Pajek format has the advantage of enabling the user to vary the sizes and shapes of the nodes. The focus in Pajek is more specifically on the visualization, while in UCINET the focus is on further analysis. Other visualization programs (e.g., NetDraw) are usually able to read one of the two formats. The user is thus able to embellish the visualizations or to pursue statistical analysis using the various options provided by these programs.

---

[2] This algorithm represents the network as a system of springs with relaxed lengths proportional to the edge length. Nodes are iteratively repositioned to minimize the overall "energy" of the spring system using a steepest descent procedure. The procedure is analogous to some forms of non-metric multi-dimensional scaling. A disadvantage of this model is that unconnected nodes may remain randomly positioned across the visualization.

I shall now discuss first the 2003 data and then the further extension in 2004, using the position of *Scientometrics* in the *Social Science Citation Index* as an example to explain the methodology. In 2004, I extend the analysis beyond *Scientometrics* and the information sciences with a focus on *Social Studies of Science* as a leading journal on the qualitative side of science and technology studies. Thereafter, I turn to "nanotechnology" in the *Science Citation Index* as an example of a policy-relevant application of the technique.

*4.1    The analysis of 2003 data*

Figure 2 provides the "being-cited" impact environment of *Scientometrics* in the *Social Science Citation Index* using Pajek for the visualization. The figure teaches us that *Scientometrics* is embedded in a group of "information science" journals, but that it also shares with *Research Evaluation* a function at the interface with *Research Policy* and *Administrative Science Quarterly*.

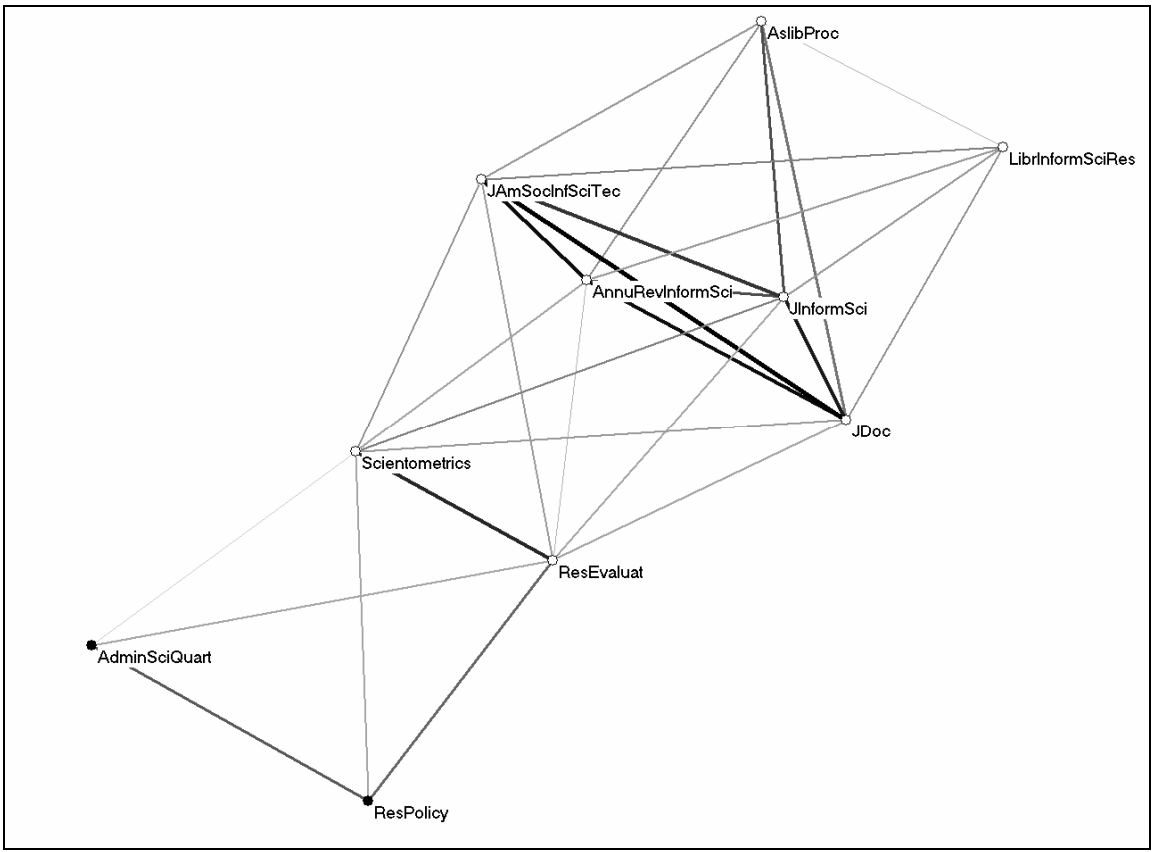

**Figure 2**: Mapping of the citation impact environment of *Scientometrics* (2003) in the cited dimension using *Pajek* for the visualization (cosine ≥ 0.2).

In Figure 3, I used NetDraw for the equivalent visualization of the journal environment, but in the "citing" dimension. From this perspective, the Croatian journal *Drustvena Istrazivanja* is added to the graph because some authors in this journal are citing according to patterns similar to those of the collective of authors in *Scientometrics* and *Research Evaluation*.[3] However, the journal was not cited in this environment by any of the other journals.

---

[3] In 2002, *Scientometrics* was not cited by *Drustvena Istrazivanja* and the citation relationship in 2003 is maintained chiefly by a single author who contributes to this journal from the perspective of scientometrics.

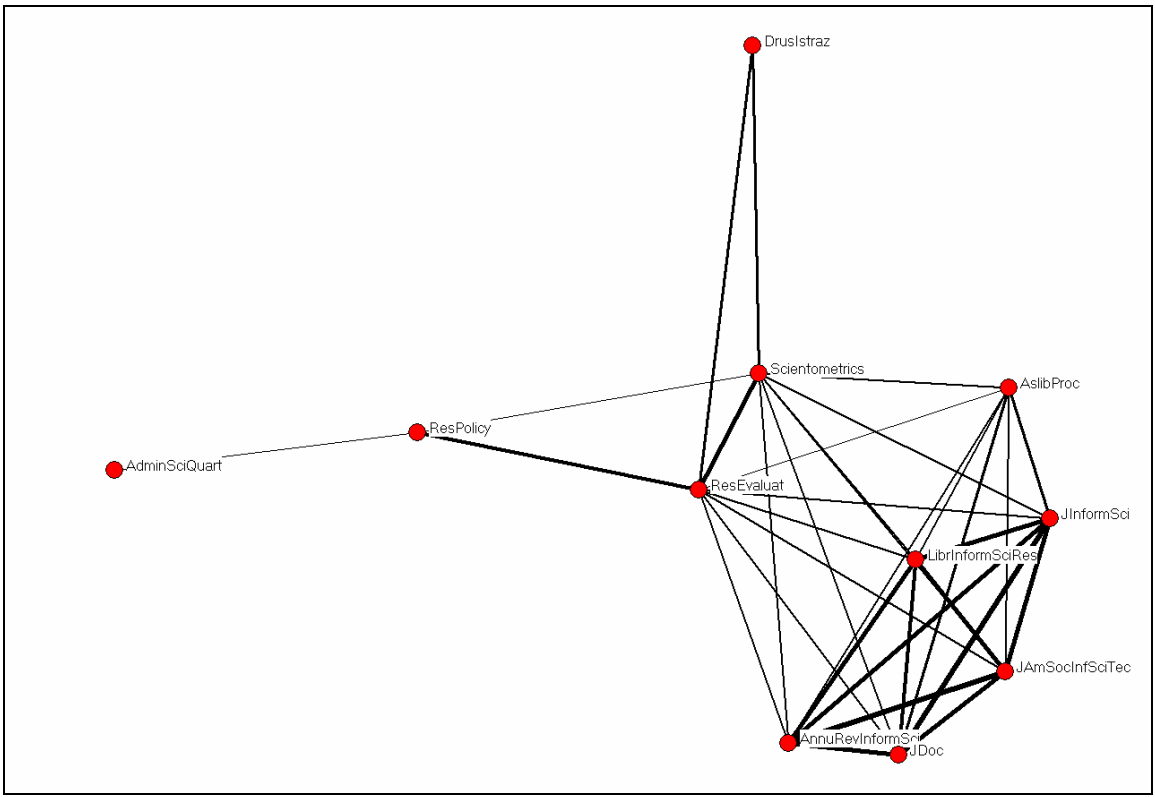

**Figure 3**: Mapping of the citation environment of *Scientometrics* (2003) in the citing dimension using *Netdraw* (cosine ≥ 0.2).

A three-factor solution of the datamatrix (explaining 73.3% of the variance) validates the conclusion of distinguishing three groups (Table 2). A first group of journals cites according to a pattern of the information sciences; a second pattern is specific for *Scientometrics* and *Research Evaluation;* and the third group can be designated as (science) policy analysis.

**Rotated Component Matrix(a)**

|  | Component | | |
|---|---|---|---|
|  | 1 | 2 | 3 |
| *J Am Soc Inf Sci Tec* | .964 | .112 |  |
| *Annu Rev Inform Sci* | .933 |  |  |
| *J Inform Sci* | .898 |  | -.236 |
| *J Doc* | .850 |  | -.220 |
| *Libr Inform Sci Res* | .791 |  |  |
| *Aslib Proc* |  |  | -.699 |
| *Res Evaluat* |  | .884 | .349 |
| *Scientometrics* | .173 | .869 | -.102 |
| *Drus Istraz* | -.325 | .640 | -.240 |
| *Res Policy* | -.216 | .371 | .750 |
| *Admin Sci Quart* | -.175 | -.301 | .565 |

Extraction Method: Principal Component Analysis. Rotation Method: Varimax with Kaiser Normalization.

a  Rotation converged in 4 iterations.

**Table 2**: Factor analysis of citing patterns of journals in the citation environment of *Scientometrics*.

*4.2    Extension of the analysis to 2004 data*

As noted above, various reactions to the 2003 version convinced us to distinguish more sharply between the citing and cited environments (Zhou & Leydesdorff, 2005). Thus, the one-percent threshold was set with reference to the "total citing" or the "total cited" of the seed journal used for the analysis. Journals can be expected to vary in terms of being relative "sinks" or "sources" of citations (Garfield, 1979).

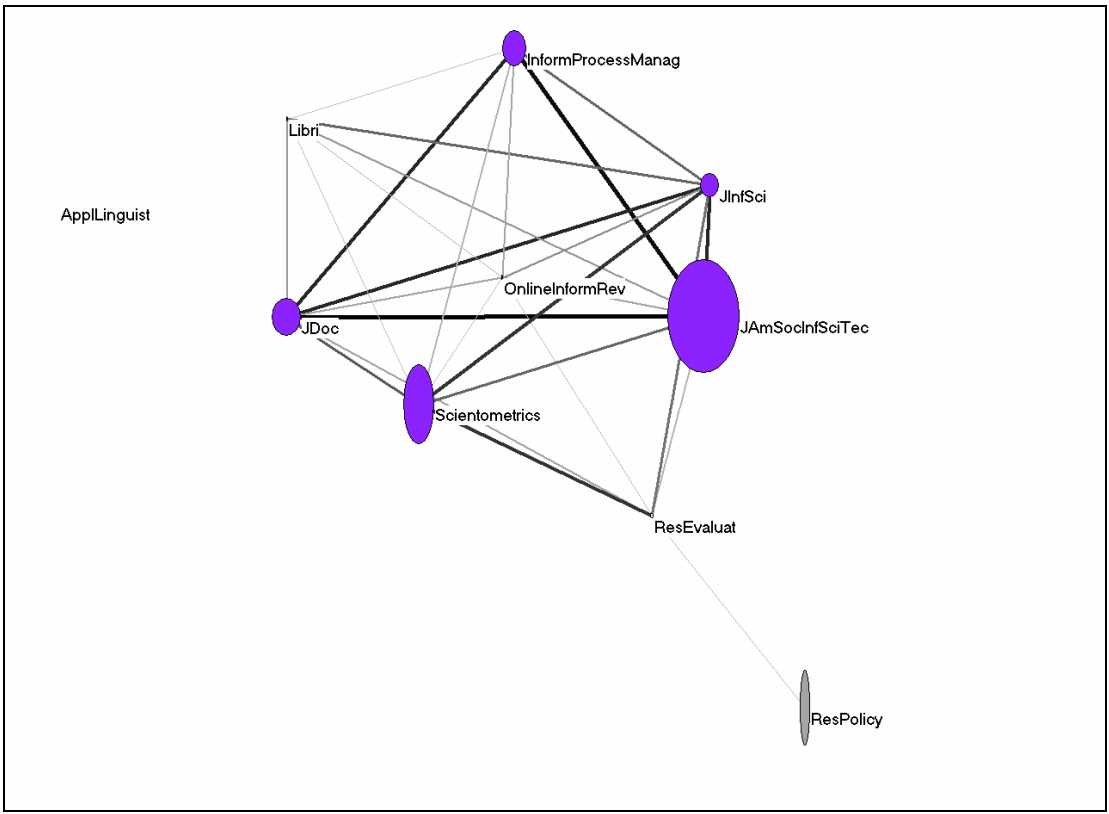

**Figure 4**: Mapping of the citation environment of *Scientometrics* (2004) in the cited dimension using Pajek (cosine $\geq 0.2$).

Figure 4 provides the citation environment of *Scientometrics* 2004 in the cited dimension using the Pajek format. From this ("being-cited") perspective, *Scientometrics* is embedded in a graph with a number of other information science journals even more strongly than in the previous year. *Research Policy* and *Applied Linguistics* are also drawn into this environment. While entertaining citation relations with the other information science journals, *Research Evaluation* functions as a bridge with *Research Policy* (cosine = 0.27). Although *Research Policy* contributes almost seven percent (55) to the citations of *Scientometrics* in 2004 (860), it has a very different pattern of citations. *Applied Linguistics* has a citation pattern completely different from any of the other journals because this journal cites some information science journals, but is not cited by any of them.

*Scientometrics* is cited 366 times (42.6%) by other papers in this same journal; only 222 citations (25.8%) are provided by the nine journals in the citation environment. The remaining 274 citations are provided by approximately 100 other journals.[4] However, the within-journal (self-)citation rate of 366 outnumbers both the other citations and the tail of the distribution. In other words, the journal *Scientometrics* is mainly cited by authors who publish in it.

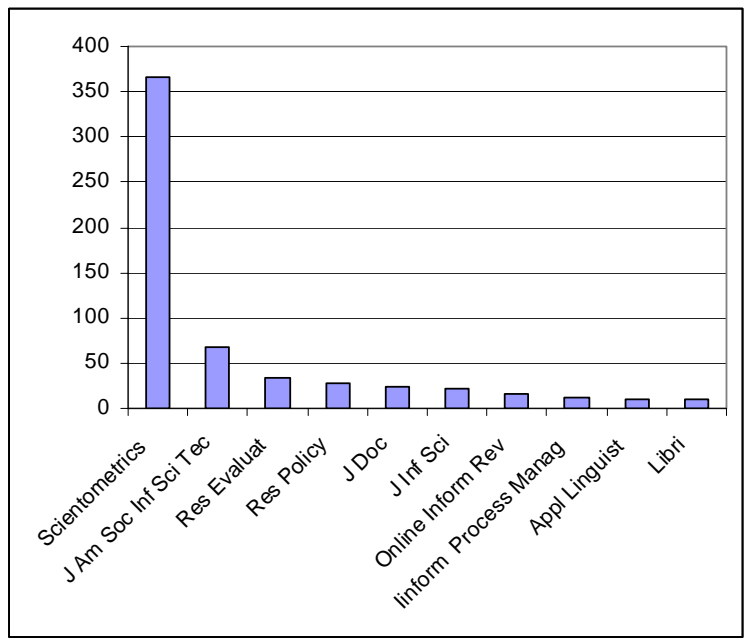

**Figure 5**: The distribution of citations provided to *Scientometrics* in 2004.

Although within-journal citations are often not self-citations by authors, they can be considered as an indicator of the inwardness of a community supporting a journal. A within-journal citation rate of more than 42% as in the case of *Scientometrics* (2004) is relatively high. The *Journal of Documentation*, for example, had in the same year only 61 (9.3%) within-journal citations of a total of being cited 654 times. Accordingly, the representation of the node for this journal in figure 4 is more like a circle, while *Scientometrics* is represented as an ellipse with a horizontal axis much shorter than the vertical one.

---

[4] Among the journals which cite *Scientometrics* above the threshold are three journals which are not processed in the *Social Science Citation Index* and therefore not included in this analysis. Two of these journals (the *Proceedings of the National Academy of Sciences of the U.S.A. (PNAS)* and the *Revista Española de Enfermedades Digestivas*) are included in the *Science Citation Index* and their relation with *Scientometrics* can be retrieved from the corresponding file in this database at http://www.leydesdorff.net/jcr04/cited . The citation rate by *PNAS* is an effect of the special issue on knowledge visualization issued by this journal in April 2004 (Shriffin & Börner, 2004).

```
*Vertices 10
1 "ApplLinguist" 0.0 0.0 0.0 x_fact 0.000000 y_fact 1.410437
2 "InformProcessManag" 0.0 0.0 0.0 x_fact 6.346968 y_fact 9.273625
3 "JAmSocInfSciTec" 0.0 0.0 0.0 x_fact 18.758815 y_fact 29.337094
4 "JDoc" 0.0 0.0 0.0 x_fact 7.616361 y_fact 9.767278
5 "JInfSci" 0.0 0.0 0.0 x_fact 4.795487 y_fact 6.346968
6 "Libri" 0.0 0.0 0.0 x_fact 0.634697 y_fact 1.163611
7 "OnlineInformRev" 0.0 0.0 0.0 x_fact 0.528914 y_fact 1.339915
8 "ResEvaluat" 0.0 0.0 0.0 x_fact 0.811001 y_fact 1.339915
9 "ResPolicy" 0.0 0.0 0.0 x_fact 2.679831 y_fact 19.358251
10 "Scientometrics" 0.0 0.0 0.0 x_fact 7.757405 y_fact 20.662906
*Matrix
0.000000 0.000000 0.000000 0.000000 0.000000 0.000000 0.000000 0.000000 0.000000 0.000000
0.000000 0.000000 0.901230 0.813610 0.567065 0.272871 0.334979 0.000000 0.000000 0.316497
0.000000 0.901230 0.000000 0.940527 0.806805 0.384551 0.328352 0.311033 0.000000 0.566747
0.000000 0.813610 0.940527 0.000000 0.826643 0.395131 0.354960 0.338886 0.000000 0.612842
0.000000 0.567065 0.806805 0.826643 0.000000 0.600559 0.455367 0.523022 0.000000 0.753862
0.000000 0.272871 0.384551 0.395131 0.600559 0.000000 0.295597 0.000000 0.000000 0.242298
0.000000 0.334979 0.328352 0.354960 0.455367 0.295597 0.000000 0.244246 0.000000 0.236972
0.000000 0.000000 0.311033 0.338886 0.523022 0.000000 0.244246 0.000000 0.274056 0.768357
0.000000 0.000000 0.000000 0.000000 0.000000 0.000000 0.000000 0.274056 0.000000 0.000000
0.000000 0.316497 0.566747 0.612842 0.753862 0.242298 0.236972 0.768357 0.000000 0.000000
```

**Table 3**:

Input file for the representation of the citation environment of *Scientometrics* 2004.

The use of the data definition language of Pajek in 2004 enables us to add parameters for the visualization. I shall also use this option for storing quantitative information about citation contributions of the respective journals. Table 3 provides the text file which corresponds to Figure 4 as an example. (This file can be retrieved at http://www.leydesdorff.net/jcr04s/cited/v1505.txt.) The ten journals in this citation environment are first defined as vertices with a label. Thereafter, three parameters are available for the coordinates of the nodes in the x, y, and z-direction (which I will not use in this study). The two parameters "x_fact" and "y_fact" provide a value for the magnification of the node in the two main directions. (Other parameters can be added, for example, in order to change the shape of the nodes from circles and ellipses into triangles, boxes or diamonds.) One should be aware that the information contained in the parameters will be lost if the Pajek-files are subsequently exported into the DL-format, for example, for the purpose of further processing in programs like NetDraw or UCINET.

In this design, I use the two parameters for indicating the percentage contribution to the thus selected citation environment both including and excluding within-journal citations. For example, *Scientometrics* was cited within this environment—that is, by these nine other journals—586 times of which 366 were within-journal citations. The total number of citations in the citation matrix among these ten journals—the grandsum ($N = \sum c_{ij}$)—is 2,863 and thus, the percentage of the citations obtained by *Scientometrics* within this environment is (586/2,836) * 100 = 20.66%. This percentage is conveniently used as the value of the parameter "y_fact." After correction for within-journal citations, the percentage become ((586-366)/2,836) * 100 = 7.76%. This value is analogously used for the parameter "x_fact." Thus, the ASCII file provides both the inputs needed for drawing the picture in Pajek and the numerical information about these percentages for those users who are interested quantitatively in the local impact factors of journals in specific citation environments.

The local impact factors are expressed as percentage shares of the grandsum of the citation environment, since the use of percentages makes the sizes independent of the citation characteristics of the specialties under study. Note that the within-journal citation rate in any year is a constant for each journal. However, the weight of this constant in each environment ($N = \sum c_{ij}$) and in the total number of citations of the journal ($\sum c_i$) varies with the environment and thus with the choice of the seed journal. In other words, the shapes and sizes of the nodes are environment-dependent.

Figure 6, for example, shows the citation impact environment of *Social Studies of Science*. Because *Scientometrics* plays a less important role in this journal's citation environment (10.0% of the citations), the weight of the relatively large within-journal citation rate is larger and after correction only a contribution of 1.1% remains as a value for the horizontal size of this node.

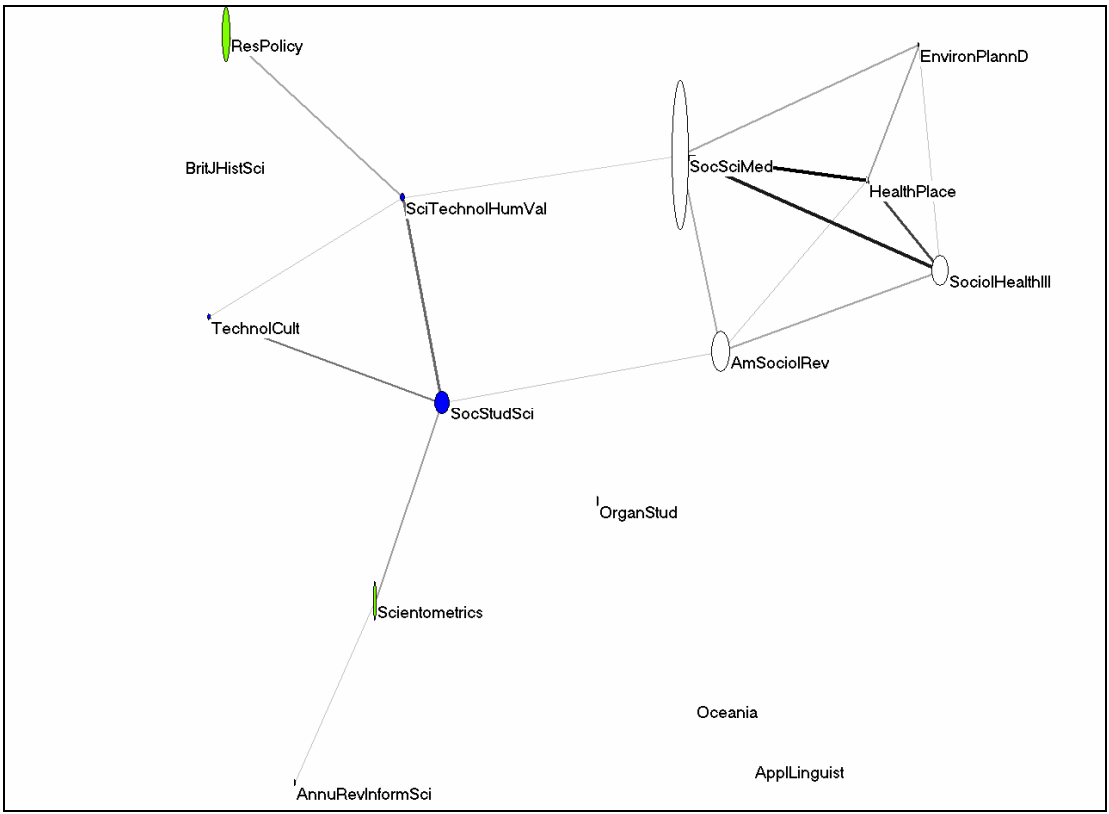

**Figure 6**: Citation impact environment of *Social Studies of Science* in 2004.

In this citation impact environment both *Research Policy* and *Scientometrics* provide relevant links, but *Research Evaluation* is no longer included. The clustering algorithm for (k-)cores in Pajek organizes *Social Studies of Science* as a separate grouping with *Science, Technology & Human Values* and *Technology and Culture*. The main environment is in this case provided by a cluster of journals in the sociology of medicine which are related also through general sociology journals like the *American Sociological Review*. *Social Science & Medicine* has the highest citation impact in this environment (37.8%) before correction for the within-journal citations; after this correction a number of journals made a contribution of between four and five percent. In terms of positions, *Scientometrics* seems the gateway to the information sciences, and *Research Policy* spans another dimension in the citation patterns among these journals. Four other journals are citing *Social Studies of Science*, but are not connected or, in other words, have a very different citation pattern.

*4.3    Citing patterns*

While being-cited patterns can be considered as impact which is largely beyond the control of the authors who are cited, citing patterns are produced by the collective of authors publishing in a certain journal in the year under study. These patterns therefore reveal how this community perceives its relevant environments at the time. Again, this perception can meaningfully be distinguished in terms of within-journal citations and citations of other journals. The two pictures of cited and citing are coupled by the within-journal citations because this number is the same in both directions.

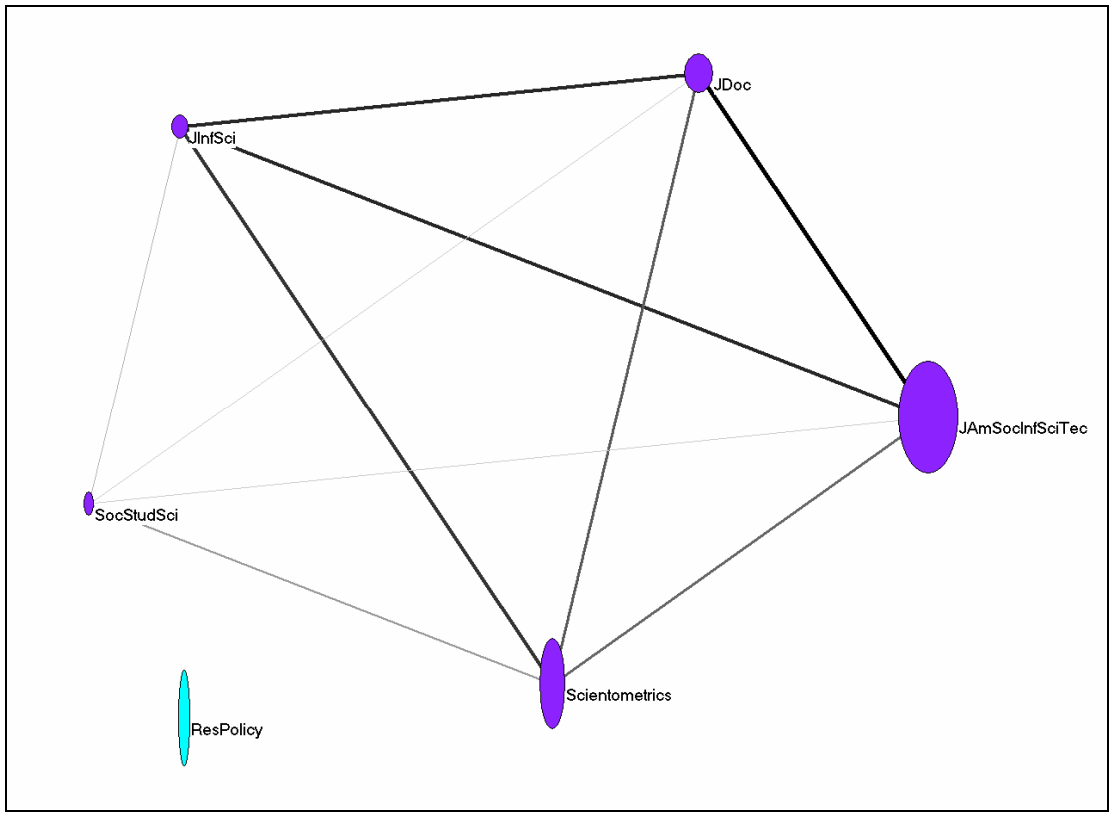

**Figure 7**: Citing patterns in *Scientometrics* 2004 (cosine ≥ 0.2)

Articles in the journal *Scientometrics* cited to a total of 2,199 times during 2004. Since the threshold of one percent thus applies to all journals cited more than twenty-one times, the tail of the distribution is much larger than in the cited dimension. The journal reaches out to a broader citing literature, but the core is at the same time more focused than in the being-cited dimension. Only five other journals are cited by the aggregate of authors in this journal above the threshold of one percent. Consistent with the results shown in Figure 6, *Social Studies of Science* is now included because it is cited by *Scientometrics* above the threshold level. The set is thus a partial subset of the journal set drawn into the citation environment in the cited direction.

The large vertical size of the ellipses representing the nodes for the *Journal of the American Society for Information Science and Technology (JASIST)* and *Research Policy* indicate that authors in these two journals are the most actively citing partners in this citation environment. The horizontal axis, however, indicates that the citation pattern of authors in *Research Policy* is dominated in this case by the within-journal citation rate (473/514 or 92%). Authors in this journal do not in this environment cite journals other than *Scientometrics* (28 times) and *Social Studies of Science* (13 times). However, in the opposite direction, *Research Policy* was cited ten times by scientometricians publishing in *JASIST*. *Research Evaluation* was not sufficiently cited by authors publishing in *Scientometrics* to pass the threshold in 2004.

When the same exercise is repeated for the citing dimension of *Social Studies of Science,* one finds no other journals included in this citation environment. The first journal cited by *Social Studies of Science* on this list after the 79 within-journal citations is *Technology and Culture* which was cited 18 times. This is below the 1% threshold because the journal cites a very large cloud of other journals for a total of 1,913 citations. Among these journals are also journals which belong to the science subjects under study in the articles. Thus, this journal has a complex citation pattern. Using this visualization method, one would have to draw only a single vertical line with zero width. In other words, this group of authors is a very specific set in terms of its citation behaviour and in such cases our methods break down.

The solution to this problem is to set the threshold for including journals at a much lower level. Figure 8 provides the citation patterns in the citing dimension for *Social Studies of Science* when the threshold is lowered to 0.1%, that is, being cited two or more times.

Thirty-eight journals are included. Note that *Scientometrics* is not among them; *JASIST* is one among six journals which are present in the environment, but not connected to any of the graphs in it (that is, above the level of cosine $\geq 0.2$).

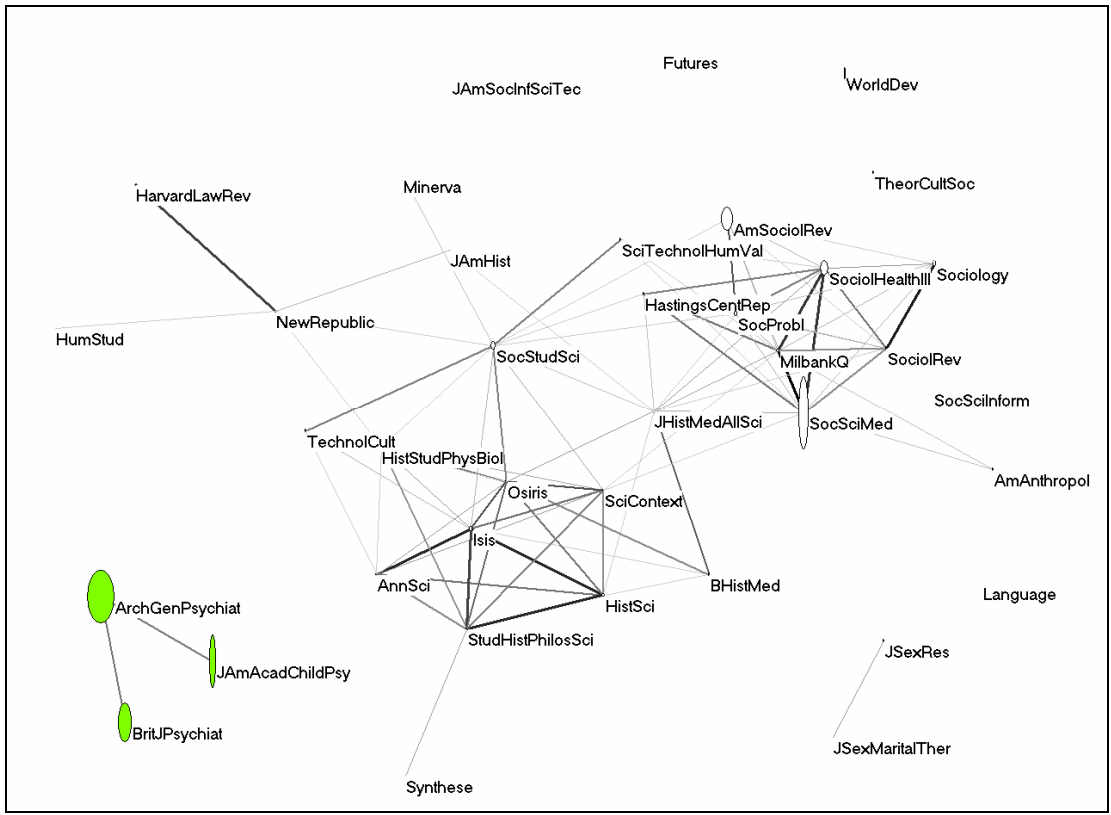

**Figure 8**: Citing environment of *Social Studies of Science* in 2004 (cosine $\geq 0.2$)

Authors in journals which contain studies about other sciences can also be expected to cite journals in these sciences. However, the citation patterns of these journals are not similar and thus the visual representations show them as isolated points or unconnected graphs (like the one with three psychiatry journals shown in Figure 8).

Some of the journals containing reflexive studies from science studies and information science are also included in the *Science Citation Index*. However, the interfaces are thin with the exception of the relations between journals in the computer sciences and information science. In 2004, one is also able to retrieve a citation relation of bibliometrics and scientometrics with the *Proceedings of the National Academy of the*

*Sciences of the USA (PNAS)* because this journal had a special issue on the subject of visualization of knowledge in which several authors used bibliometric techniques (Shriffin & Börner, 2004).

## 5. Policy relevance: Nanotechnology

In their methodological annex to one of the first systematic evaluations of science policy priorities, Studer & Chubin (1980, p. 269) noted the need to construct a baseline against which to measure the effectiveness of science policy programs. Can journals and their relations be used as such a baseline? On the basis of an analysis of the effects of some major breakthroughs in the sciences during the 1980s, Leydesdorff *et al*. (1994) concluded that the new developments in the sciences can be expected to lead to an expansion of the literature in the relevant domains, and that, structural changes can first be noted in the "cited" dimension of the journal-journal citation matrix. The new journals are cited in the neighbouring disciplines, while the citation patterns within the new journals still have to crystallize as a structure among them. Authors publishing in these journals position their papers with reference to relevant environments and thus trigger a process of coevolution with the already existing structures.

In other words, "interdisciplinarity" in this sense means that authors (1) draw upon existing disciplines and their corresponding journal structures, and (2) achieve visibility in publications in previously existing disciplines by being cited, but that (3) it is still an open question whether this development will lead to new discipline formation in the years thereafter. Let us apply this reasoning to a current development using the tool introduced above, taking nanotechnology as an example.

Nanotechnology is a priority area in various nations. The newly emerging (inter-)discipline has been difficult to delineate (PCAST, 2005; Kostoff, 2004). Zhou & Leydesdorff (2006) distinguished between a core-set of three journals of nanotechnology and a nano-relevant set of 85 journals. The three core journals are listed in Table 4 with their respective impact factors and total number of cites as the two main indicators of impact as distinguished above. On both indicators *Nano Letters* is ranked as the leading journal.

| Journal titles | Impact factor | Total cites |
|---|---|---|
| *Journal of Nanoscience and Nanotechnology* | 2.017 | 489 |
| *Nano Letters* | 8.449 | 7349 |
| *Nanotechnology* | 3.332 | 2858 |

**Table 4**: Three core journals in nano-technology with their impact factors and their total numbers of being-cited in 2004 (Source: *JCR-SCI*)

*Nano Letters* is published by the American Chemical Society and can therefore be considered as a central journal also at the level of the database (Bensman, 2001; Leydesdorff & Bensman, forthcoming). Let us import the cited and the citing patterns of this journal into Pajek as a kind of experiment. (The respective input files for the next two figures are provided at http://www.leydesdorff.net/jcr04/cited/v4399.txt and http://www.leydesdorff.net/jcr04/citing/v4399.txt.)

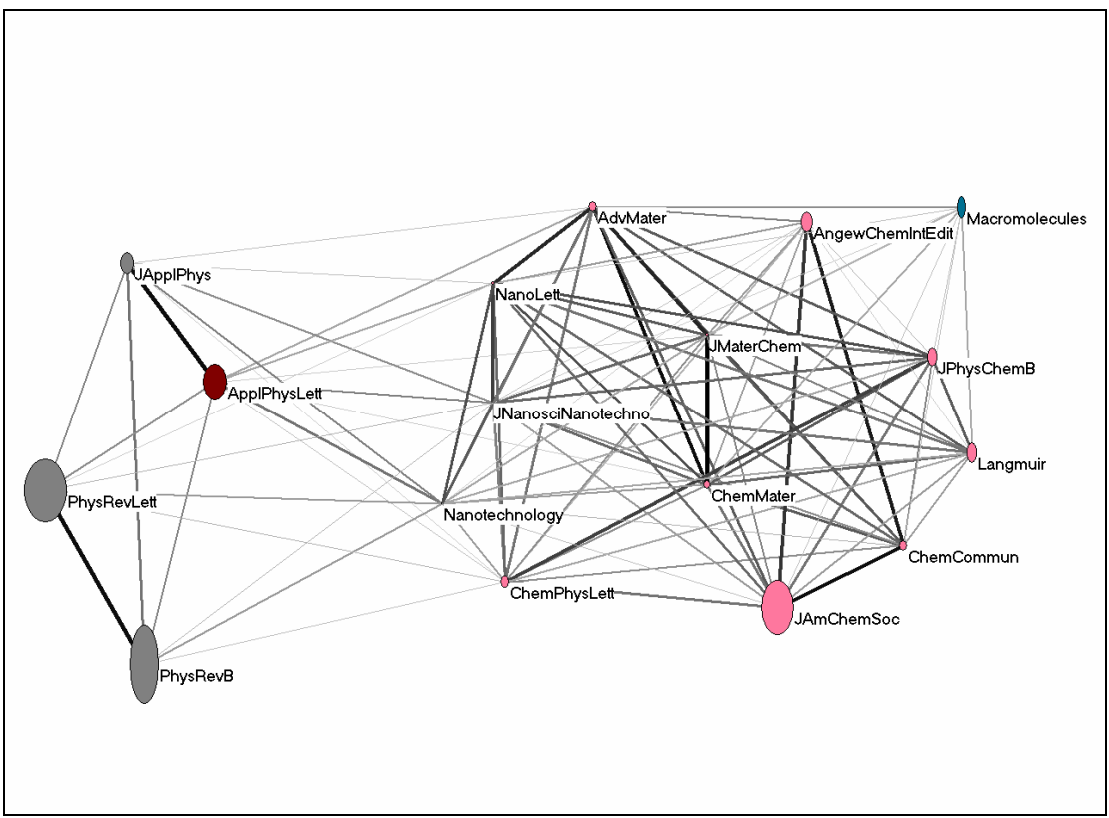

**Figure 9**:

Citation impact environment ("cited") of *Nano Letters*; 17 journals; cosine $\geq$ 0.2.

Figure 9 shows the citation impact environment of *Nano Letters.* The journal is relevant to (1) the other core journals of nanotechnology distinguished above, (2) the major journals of chemistry via the interface of chemical materials, and (3) physics through the interface of applied physics. Some smaller journals in chemical physics and material sciences are relevant in this environment as well, as are a few journals in polymer chemistry. The citation relation with the latter, however, is more indirect.

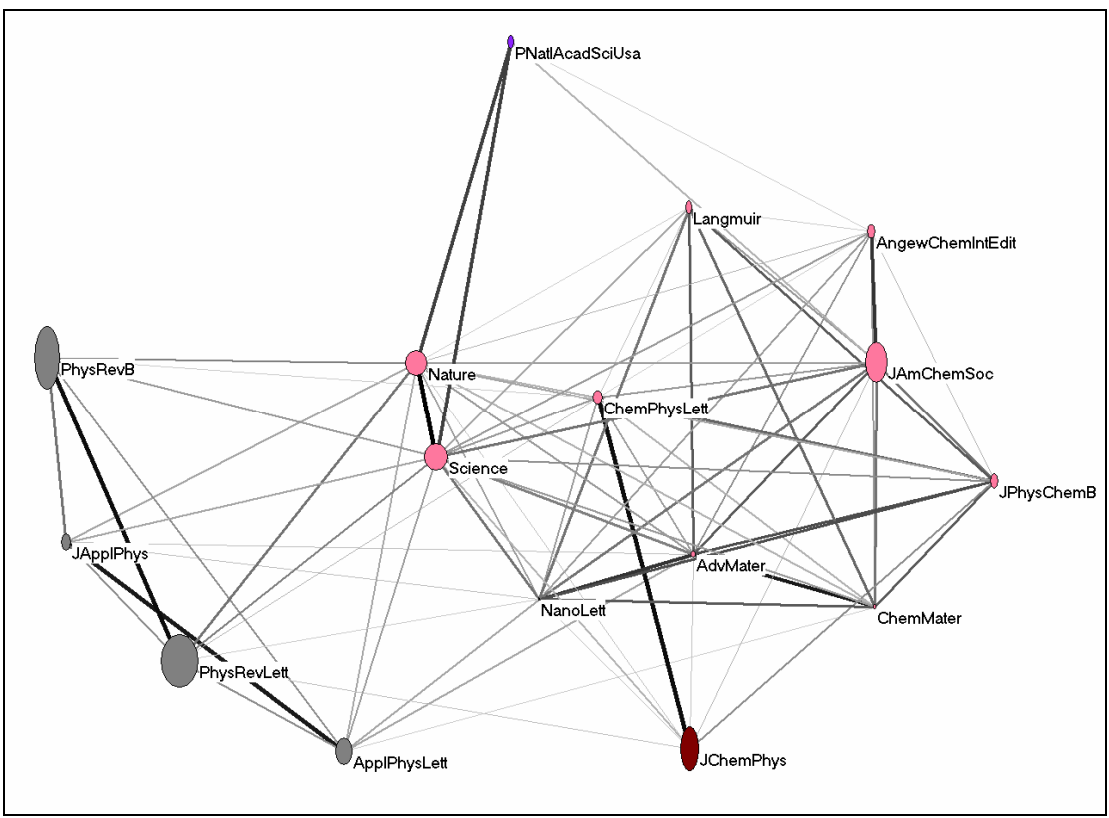

**Figure 10**: The citing environment of the journal *Nano Letters*; 16 journals; cosine ≥ 0.2

Figure 10 shows us that articles in *Nano Letters* do not cite papers in the other core journals of nanotechnology above the one percent threshold, but rather major journals in general science like *Nature*, *Science, PNAS*, and (sometimes leading) journals in the neighbouring disciplines. A similar picture can be obtained using the other core journals of nanotechnology distinguished above. Thus, nanotechnology journals have not yet formed a core set which cite each other actively, but by "being cited" they are in each other's environments (because of the lower absolute values of the thresholds in this dimension). Articles in these journals, however, provide large numbers of references to journals in more established fields, and the prominent inclusion of major science journals in this environment suggests that this is also done for purposes of legitimation.

In other words, these results suggest that new developments at interdisciplinary interfaces in the sciences first generate an effect on the relevant citation environments and can thus be made visible in the cited environments. In the citing environments, authors in these journals cannot yet sufficiently legitimize their results in terms of the other journals within the cluster because the specialty is not yet sufficiently stabilized and codified. Therefore, one can expect leading journals to be cited frequently and prominently. In a later stage of development, the specialist journals may turn towards one another and form a strongly connected bi-directional graph.[5]

## 6. Conclusions and discussion

The normalization of any impact measure requires the delineation of a relevant set (Pudovkin & Garfield, 2002). The ISI journal set is by definition a "mixed bag" and its classification into subsets is therefore dependent on the choice of a perspective. Given a perspective, the delineations remain additionally sensitive to the inclusion or exclusion of (e.g., interdisciplinary) journals in the relevant environments. Recently, it has become clear that for theoretical reasons no single "best" classification is possible (Bensman, 2001; Leydesdorff & Bensman, forthcoming). The choice in this study of a journal-based organization of relevant environments is based upon an awareness that the aggregation is never robust and therefore is best left to the user. This does not mean that any aggregation will do—like the seemingly arbitrary or at least insufficiently legitimated classification of the ISI, which is nevertheless used intensively (Leydesdorff, forthcoming)—but that one needs substantive or methodological arguments for the aggregation. Ideally, a system should allow users to select one or more journals online and then generate the relevant environments and corresponding matrices dynamically. I have developed such a system offline, but I cannot bring it online both for technical reasons and because of copyright restrictions.[6]

Nevertheless, the availability of the input files for Pajek enable the user to gain insights into the local citation environments of journals, and by choosing different seed journals one can obtain a conclusive answer to practical questions. I have shown this above extensively for the reflexive journals *Scientometrics* and *Social Studies of Science.* The focus enabled me to interpret the complex patterns of citations among journals in the social sciences and discuss methodological problems in a comprehensive framework and considerable detail. The citation matrix of the *Science Citation Index* is more dense and profiled than that of the *Social Science Citation Index*, and therefore easier to analyze in principle. *Nano Letters* provides a good example of how one can access a policy priority area in terms of both its being-cited or impact dimension and its citing or activity dimension.

---

[5] Van den Besselaar & Leydesdorff (1996) provide a detailed analysis of such a turn for the case of journals in artificial intelligence around 1988.
[6] The envisaged procedure would require that the copyrighted data were brought on the server. In the current procedure, however, this is not the case: only the results of the computation are brought online.

In a further extension I followed Bensman's (forthcoming) argument for total citations in a delineated set as an indicator of the prestige of a journal rather than its short-term impact. Price (1965) already wrestled at the beginning of his research on journal relations with the two types of scientific impact that can roughly be distinguished as the current impact of articles at the research front and the longer-term impact of reviews with archival functions. Review articles can be expected to retain the knowledge obtained in the recent period in a next layer of codification. However, it was a surprise for me to find how these two dimensions can differ so sharply (Figure 1). Bensman's study shows that faculty identifies with prestige and not with impact as defined by the ISI.

An additional correction for within-journal citations was needed when we explored this technique further—in the time between developing the 2003 and 2004 versions—using the journal set of the Institute of Scientific and Technical Information of China (ISTIC). In the Chinese data, we found a high degree of unevenness in the cited/citing ratios and sometimes up to 100% within-journal citation rates (Leydesdorff & Jin, 2005; Zhou & Leydesdorff, 2005). The appreciation of the outliers on the main diagonal as a separate source of information solves an old problem in scientometrics about the appropriate normalization of these cells (Price, 1981; Noma, 1982): One can appreciate the within-journal citations as additional information about the journal structure.

In the above exercises, the cutoff level of 1% of the total cited or total citing rate, respectively, has remained insufficiently discussed. I showed that in the case of *Social*

*Studies of Science* reducing this cutoff by an order of magnitude to 0.1% provided a means to visualize the citing environment despite the extremely sparse network of citations other than within-journal citations (Figure 8). He & Pao (1986, p. 410) noted the difference in the skewness of the distributions in the citing and cited directions, and therefore used a 1% cutoff for "citing" and a 2% cutoff for "cited." If necessary, I can bring the data online in future years using a lower cutoff level. However, ideally the cutoff levels should be determined by the shapes of the distributions. The latter can be expected to vary among fields of science and perhaps to a lesser extent among journals within fields, but they all contain a large tail which is loglinear and a core set which is curvilinear. Elsewhere, we have shown that the curvilinear part is the relevant one, and a formal criterion would thus be needed to distinguish this part from the other (Leydesdorff & Bensman, forthcoming). Such a criterion should first be tested across the file, and for both the *Science Citation Index* and the *Social Science Citation Index,* because the two indices tend to differ in terms of the relevant distributions (Price, 1970; Nederhof *et al.*, 1989; Van der Meulen & Leydesdorff, 1990).

**Acknowledgement**
The author is grateful to Stephen J. Bensman for providing relevant data and to Ping Zhou for complementing this dataset.